\documentclass[twocolumn,superscriptaddress,amsmath,amssymb,showpacs]{revtex4}
\usepackage{amssymb}
\usepackage{amsmath}
\usepackage{dcolumn}
\usepackage{bm}
\usepackage{graphicx}
\usepackage{amsfonts}

\begin{document}

\title{Three dimensional cooling and detecting of a nanosphere with a single cavity}
\author{Zhang-qi Yin}
\email{yinzhangqi@gmail.com}
\affiliation {State Key Laboratory of Magnetic Resonance and Atomic
and Molecular Physics, Wuhan Institute of Physics and Mathematics,
Chinese Academy of Sciences, Wuhan 430071, China }
\affiliation{Key Laboratory of Quantum Information, University of
Science and Technology of China, Chinese Academy of Sciences, Hefei, Anhui 230026, China}
\author{Tongcang Li}
\affiliation{Center for Nonlinear Dynamics and Department of
Physics, The University of Texas at Austin, Austin, TX 78712, USA}
\author{M. Feng}
\email{mangfeng@wipm.ac.cn}
\affiliation {State Key Laboratory of Magnetic Resonance and Atomic
and Molecular Physics, Wuhan Institute of Physics and Mathematics,
Chinese Academy of Sciences, Wuhan 430071, China }

\begin{abstract}
We propose an experimental
scheme to cool and measure the three-dimensional (3D) motion of an
optically trapped nanosphere in a cavity. Driven by three lasers on
TEM00, TEM01, and TEM10 modes, a single cavity can cool a trapped
nanosphere to the quantum ground states in all three dimensions
under the resolved-sideband condition. Our scheme can also detect an
individual collision between a single molecule and a cooled
nanosphere efficiently. Such ability can be used to measure the mass
of molecules and the surface temperature of the nanosphere. We also
discuss the heating induced by the intensity fluctuation, pointing
instability, and the phase noise of lasers, and justify the
feasibility of our scheme under current experimental conditions.

\end{abstract}
\pacs{42.50.Wk, 37.10.Vz, 05.40.Jc, 42.50.Ct}
 \maketitle
\section{introduction}

Cooling microscopic, mesoscopic, and macroscopic objects to their
motional ground states has attracted great attention in the past
decades. Various atoms, ions and molecules have been cooled and
trapped, and some of them have been employed in quantum information
processing and atomic clocks. It is of fundamental interests to cool
macroscopic objects down to quantum regime for studying quantum
effects in macroscopic systems, improving precisions in
ultra-sensitive measurements \cite{Clerk2010,2010arXiv1003.3752A,PhysRevLett.104.133602},
and realizing quantum information processing with new ideas
\cite{PhysRevLett.90.137901,PhysRevLett.102.020501,YH09}. Cooling
mechanical oscillators near the ground state can be
accomplished by placing the high frequency oscillator in cryogenic environment \cite{LaHaye2004,Connell2010},
or optomechanical cavity cooling methods \cite{2006Natur.443..193N,2006Natur.444...67G,Schliesser2008,2007PhRvL..99i3902M,2007PhRvL..99i3901W},
or combining them together \cite{Schliesser2009,2009NatPh...5..485G,2009NatPh...5..489P,2010Natur.463...72R}.

Recent report has shown the possibility to cool a mesoscopic
microwave-frequency mechanical oscillator down to the motional
ground state by standard cryogenic methods \cite{Connell2010}.
However, the mechanical Q factor (around 260)  in this system
\cite{Connell2010} is too small for many applications. Similar to
optical trapping and cooling of atoms \cite{PhysRevLett.84.3787,PhysRevA.64.033405,PhysRevLett.103.103001} and molecules
\cite{PhysRevA.81.063820,PhysRevA.77.023402}, a nanosphere
can be optically trapped and cooled in a cavity
\cite{1976ApPhL..28..333A,Chang2010,Romero2010,2010arXiv1010.3109R}. An optically
trapped nanosphere in vacuum is well isolated from the thermal
environment and can have a mechanical Q factor larger than
$10^{10}$. This approach has the potential to cool a mechanical
system to the vibrational ground state even at room temperature,
based on which nonclassical states(e.g. squeezed states) could be
generated. A cooled nanosphere can also be used to test gravity
induced decoherence effects \cite{penrose96} and search for
non-Newtonian gravity forces \cite{PhysRevLett.105.101101}.

We  noticed that the first part of the proposal
\cite{Chang2010,Romero2010}, which is trapping micro(nano)-sphere by
optical tweezer with high frequency, has been realized
experimentally \cite{Litongcang2010}, in which a glass microsphere
was optically trapped in air and vacuum, and its Brownian motion was
measured with ultrahigh precision. A more exciting work would be to
cool a nanosphere to the quantum ground state using sideband cooling
with the help of cavities \cite{Chang2010,Romero2010}, and observe
the individual collisions between the sphere and single molecules
\cite{PhysRev.23.710}.
 A nanosphere will scatter the cooling laser to all three
dimensions and cause 3D heating. The heating effects of laser noises
are also 3D. As will be discussed later, such heating can cause
exponential growth of the kinetic energy of a nanosphere. If only
one-dimensional motion is cooled efficiently, the others will be
heated up continuously and the nanosphere will be kicked out of the
trap. In order to achieve ground state cooling of an optically
trapped nanosphere, we must use a 3D cooling scheme. We can
straightforwardly add two more cavities for cooling the other two
dimensions, but the system will become too complex to be realized.
We may combine the 1D cavity cooling with 2D feedback cooling to
stabilized the system. But the system will also become complex and
can only do ground state cooling in 1D.

In this work, we propose to cool and measure the 3D motion of a
nanosphere by TEM00, TEM01, and TEM10 modes of a single cavity. We
show that each one of these three modes can be coupled to the motion
of a trapped nanosphere in each dimension respectively. Thus they
can be used to cool and detect the 3D motion of a nanosphere. The
scheme can be used for detecting the individual collisions between
molecules and the nanosphere. The mass of the molecules, and the
surface temperature of the nanosphere may also be measured at the
same time. We noticed trapping  single atoms in a  high-finesse
cavity  driven by three lasers at TEM00, TEM01, and TEM10 modes
simultaneously has been realized in an experiment
\cite{PhysRevLett.99.013002}. One can also use a phase plate to
generate a TEM01 (or TEM10) beam from a TEM00 beam \cite{Meyrath05},
and use it to pump the corresponding mode of a cavity. Our scheme
should also helpful for cavity cooling of atoms (ions) and
molecules.

The paper is organized as follows. In Sec. \ref{sec:3D} , we introduce the scheme of
nanosphere 3D cooling via a cavity. In Sec. \ref{sec:detecting}, we propose the scheme of
detecting the collisions between molecules and the sphere, and discuss
the experimental possibility. In Sec. \ref{sec:c}, we give a short summary of the paper.


\section{3D cooling model} \label{sec:3D}

\begin{figure}[htbp]
\centering
\includegraphics[width=7cm]{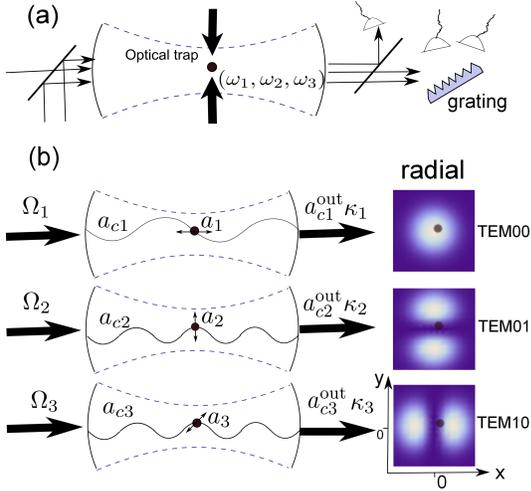}
\caption{(color online) (a) Cooling and detecting scheme. A
nanosphere is trapped by a dual-beam optical tweezer inside of a
cavity. The cavity is driven by three lasers in TEM00, TEM01 and
TEM10 modes. The TEM01 mode laser has different polarization, and is
separated from the other two lasers by a polarizing beam splitter
for detection. The TEM00 and TEM01 lasers have different
frequencies, and are separated by a grating for detection. (b) Three
cooling modes TEM00, TEM01, and TEM10, and their radial
distribution. The black dot represents the position of a trapped
nanosphere.} \label{fig:scheme}
\end{figure}

As shown in Fig. 1a, we consider an optically trapped nanosphere
with mass $m$ confined in a cavity by means of an optical tweezer
\cite{Litongcang2010}. Since the mechanical Q of the system could be
extremely high, e.g., $> 10^{10}$ \cite{Chang2010,Romero2010}, we
may consider an ideal system in the first part of our treatment, but
leave the effect from the environment, such as the collisions
between molecule and nanosphere, to later discussion. The
frequencies of the optical trap along the $z$, $x$, and $y$ axes are
$(\omega_1$, $\omega_2$, and $\omega_3)$. Contrary to the
conventional method of using a cooling laser with TEM00 mode to cool
the motion along $z$ direction, we add two non-Gaussian beams with
TEM01 and TEM10 modes to drive the cavity in order to cool the
motion along the $x$ and $y$ directions, respectively. The resonant
frequencies of the cavity modes $a_{c1}$, $a_{c2}$, and $a_{c3}$ are
$\omega_{c1}$, $\omega_{c2}$, and $\omega_{c3}$, respectively. The
detunings between the lasers and the cavity modes are $\Delta_{cj} =
\omega_c^j - \omega_L^j$ $(j=1,2,3)$. We suppose that the TEM01 and
TEM10 lasers have the same frequency, but with orthogonal
polarization. The TEM00 and TEM01 lasers have the same polarization,
but different frequencies. In practical, the frequency differences between TEM00 and
TEM01 (TEM10) could be very large, and the TEM01 and TEM10 modes are orthogonal in
polarizations. Therefore the interference between the three cavity modes
can be neglected.

Supposing the radius of the nanosphere to be much smaller than the
wavelength of the cavity mode, we may calculate the sphere-induced
cavity frequency shift $\delta\omega$ by perturbation theory,
 \begin{equation}
   \label{eq:shift}
   \frac{\delta \omega}{\omega_0} = -\frac{1}{2} \frac{\int d^3 {\bf r}
    \delta P({\bf r}) \dot {\bf E}({\bf r}) }{\int d^3{\bf r} \epsilon_0
    {\bf E}^2({\bf r})},
 \end{equation}
where $\omega_0$ is the resonant frequency of a cavity without the
nanosphere, ${\bf E}({\bf r})$ is the cavity mode profile and
$\delta P({\bf r})$ is the variation in permittivity induced by the
nanosphere. Due to the tiny scale of the nanosphere, we have $P({\bf
r}')\simeq \alpha_{\mathrm{ind}}E({\bf r})\delta ( {\bf r} - {\bf
r}')$, with ${\bf r}$ the center-of-mass position of the nanosphere,
$\alpha_{\mathrm{ind}} = 3 \epsilon_0
V(\frac{\epsilon-1}{\epsilon+2})$ the polarizability, and $V$ the
sphere volume.

The total Hamiltonian of the system in the rotating frame is
\begin{equation}\label{Hamiltonian}
  \begin{aligned}
    H = &\sum_{j=1}^3 \big[\hbar \omega_j a_j^\dagger a_j -
     \hbar (\Delta_j-U_j) a^\dagger_{cj} a_{cj} +
   \frac{\hbar\Omega_j}{2} ( a_{cj} +a_{cj}^\dagger)\big],
  \end{aligned}
\end{equation}
where $a_j$ characterizes the phonon mode along $q_j$ direction with
$q_1=z, q_2=x, q_3=y$. $\Omega_j$ is the driving strength by the
lasers and $U_j$ characterizes the coupling between the cavity mode
$a_{cj}$ and the nanosphere.
 In the limit that
$\epsilon \gg 1$, where $\epsilon$ is the electric permittivity of
the nanosphere, we get \cite{Chang2010}
\begin{equation*}
   \begin{aligned}
    U_1=& -  \frac{3V}{2V_{c1}}
\exp (-\frac{2x^2+ 2y^2}{w^2}) \cos^2 (k_1z +\varphi_1)
\omega_{c1},\\
    U_2 = &-  \frac{3V}{2V_{c2}}
\frac{x^2}{w^2}\exp (-\frac{2x^2+ 2y^2}{w^2}) \cos^2 (k_2z +
\varphi_2)\omega_{c2},\\
    U_3 = &- \frac{3V}{2V_{c3}} 
\frac{y^2}{w^2}\exp (-\frac{2x^2+ 2y^2}{w^2})\cos^2 (k_3z +
\varphi_3)\omega_{c3},
   \end{aligned}
\end{equation*}
with $V_{c1}= (\pi/4)Lw^2$ and $V_{c2}=V_{c3}=(\pi/16)Lw^2$.

We assume the optical tweezer to be much stronger than the
cavity-mode-induced trap, and neglect the effects of cooling lights
on trapping. Besides, if we carefully choose the location of the
trap, such as $z_0=0$, $x_0=y_0=0.25w$, $\varphi_1=\pi/4$, and
$\varphi_2=\varphi_3=0$, the gradients of the three light fields lie
approximately along the three axes. The effective Hamiltonian is
\begin{equation}
 \begin{aligned}
 \label{eq:effH}
 H_{eff} = &  \sum_{j=1}^{3} \big[\hbar \omega_j
a_j^\dagger a_j -\hbar \Delta_j a^\dagger_{cj} a_{cj} +
\frac{\hbar\Omega_j}{2} ( a_{cj} +a_{cj}^\dagger) \\ &+ \hbar g_j
a^\dagger_{cj} a_{cj} (a_j+ a_j^\dagger) \big],
 \end{aligned}
\end{equation}
where $g_j=q_{\mathrm{zpf}j} \partial U(x,y,z)/\partial
j|_{x=x_0,y=y_0,z=z_0}$ characterizes the coupling strength between
the cavity  mode and the oscillation of the nanosphere, and
$q_{\mathrm{zpf}j}= \sqrt{\hbar/2m\omega_j}$ is zero-point
fluctuation for the phonon mode $a_j$.  In general, $g_1$ can be one
to two orders larger than $g_2$ and $g_3$. The effective Hamiltonian \eqref{eq:effH}
is deduced with linearization, which is valid when the vibration amplitude
of a trapped nanosphere is much smaller than the wavelength of the laser.
The rms vibration amplitude of a particle in a harmonic trap is  $\sqrt{k_B T
/(m\omega^2)}$ .
For a nanosphere with radius of $50$ nm trapped in an optical tweezer
with trapping frequency of $0.5$ MHz, the vibration amplitude is $20$ nm
at $300$ K, and will be only $1.2$ nm at $1$K, which are very small. Thus
the linearization will be valid if the nanosphere is pre-cooled by
feedback cooling.


 From Eq. \eqref{eq:effH}, the linearized Heisenberg equations of motion for
our system are,
\begin{equation}
  \label{eq:heisenberg}
 \begin{aligned}
  \dot{a_{cj}}=& (i\Delta_{cj}' - \kappa_j/2)a_{cj} - ig_j \alpha_j
                  (a_j + a_j^\dagger) +\sqrt{\kappa_j} a_{cj}^{\mathrm{in}},\\
 \dot{a_j} = &-i\omega_j  a_j - i g_j (\alpha_j a_{cj}^{\dagger} + \alpha_j^* a_{cj}),
 \end{aligned}
\end{equation}
where $\alpha_j = i\Omega_j/(2i\Delta_{cj}' -\kappa_j)$,
$\Delta_{cj}' = \Delta_{cj} + 2g_j^2|\alpha_j|^2/\omega_{cj}$, and $\kappa_j$ is the decay rate of the
cavity mode $a_{cj}$.
$\alpha_j$ is the amplitude of cavity mode $a_{cj}$. $\Delta'_{cj}$
is the effective detuning between the driving laser and the cavity
mode $a_{cj}$. The linearization of the Heisenberg equations is valid only
if the state is stable. The stable criteria is \cite{PhysRevA.77.033804}
\begin{equation}
  \label{eq:stable}
 \begin{aligned}
  S_1^j =& 4\Delta_{cj}' \omega_m g_j^2 \alpha_j^2 \kappa_j^2>0\\
 S_2^j = &\omega_m \Delta_{cj}{'^2}-g_j^2\alpha_j^2 \Delta_{cj}'>0,
 \end{aligned}
\end{equation}
Because of $\Delta_{cj}'>0$, the criteria $S_1^j$ are always valid. The criteria $S_2^j$
are valid only when $g_j \alpha_j <\sqrt{\omega_m \Delta_{cj}'}$.
In the following discussion, we suppose that the stable criteria \eqref{eq:stable}
is satisfied.

 To realize resolved sideband cooling, we require
$\omega_j \gg \kappa_j$. We suppose $ |g_j \alpha_j| \ll \kappa_j$, and
find that the final phonon number is \cite{Clerk2010}
$$
n_{mj}= -\frac{(\omega_j+\Delta'_{cj})^2 +(\kappa_j/2)^2}{4\omega_j
\Delta'_{cj}}.
$$
In the special case of $\Delta_{cj}= -\omega'_j$, the final phonon
number is $n_{mj} =(\kappa_j/4\omega_j)^2 \ll 1$. The cooling rate is
$\Gamma_{j} =  g_j^2 |\alpha_j|^2 /[
\kappa(1+\frac{\kappa_j^2}{16\omega_j^2})]$.

\section{Detecting scheme and noises of the scheme} \label{sec:detecting}

 The scheme can measure the 3D motion of the nanosphere at the same time.
 We have a reduced equation
under rotating wave approximation, in the case of $\Delta_{cj}' =
-\omega_j$ and $\omega_j \gg \kappa_j, \alpha_j g_j$, as \cite{PhysRevLett.98.030405,Yin2009},
\begin{equation}
  \label{eq:heisenberg1}
   \begin{aligned}
  \dot{a_{cj}}=&  - \frac{\kappa_j}{2}a_{cj} - ig_j \alpha_j
                  a_j  +\sqrt{\kappa_j} a_{cj}^{\mathrm{in}},\\
 \dot{a_j} = &- i g_j \alpha_j a_{cj}.
 \end{aligned}
\end{equation}
In the limit $\kappa_j \gg g\alpha_j$, using boundary condition
$a_{cj}^{\mathrm{out}} = -a_{cj}^{\mathrm{in}} + \sqrt{\kappa_j}
a_{cj}$, we get $a_{cj}^{\mathrm{out}} = -i
\frac{2g\alpha_j}{\sqrt{\kappa_j}} a_j + a_{cj}^{\mathrm{in}}$,
$\dot{a}_j = -\frac{2g_j^2 \alpha_j^2}{\kappa} a_j
-\frac{2ig_j\alpha_j }{\sqrt{\kappa}} a_{cj}^{\mathrm{in}}$.
Therefore the 3D motion of the nanosphere can be measured by
detecting the output fields. In the resolved sideband limit, the
output field is nearly vacuum, and will have a signal when there are
collisions between the residual molecules in vacuum and the
nanosphere. Besides, the shot noise can also be neglected in the
scheme as it is very small (estimated to be around $10^{-4}$ Hz in
Ref \cite{Chang2010}).

Because a collision between a molecule and a nanosphere is 3D in
nature, our 3D scheme will be essential for efficient detecting of
the collisions. Detection of individual collisions between single
molecules and the nanosphere would lead to a test of the
Maxwell-Boltzmann distribution on single-collision level.
Considering the gas pressure $P$ at temperature $T_\mathrm{env}$,
the radius of the sphere $r$, the molecule mass $m_m$, we have the
collision number per second  $N= (2\pi r^2)P/\sqrt{\pi m_m k_B
T_\mathrm{env}/2}$ \cite{PhysRev.23.710}, where $k_B$ is the
Boltzmann constant. The collision time is estimated to be much less
than the nanosphere oscillation time scale. The three phonon modes
initially in vacuum will be in a state with mean phonon number
$n_{j0}$: $\langle a^\dagger_j(t_0) a_j (t_0)\rangle= n_{j0}$ after
a single collision, where $t_0$ is the time when collision happens.
For this case, the output field is
$$a_{cj}^{\mathrm{out}}(t) = -i \frac{2g\alpha_j}{\sqrt{\kappa_j}} \exp
[-\frac{2g_j^2 |\alpha_j|^2}{\kappa_j}(t-t_0)]a_j(t_0) +
a_{cj}^{\mathrm{in}},$$
 It is easy to
find that $\int_{t_0}^\infty\langle a_{cj}^{\mathrm{out}}(t)
a_{cj}^{\mathrm{out}\dagger}(t)\rangle dt =n_{j0}$. This implies
that the output-pulse photon number is equal to the increase of the
phonon number after the collision. From above discussion, we get the
phonon decay time $\tau_j = \kappa_j/(4g^2_j |\alpha_j|^2)$, which
is also the pulse duration of the output light of mode $a_{cj}$. The
phonon number can be measured by detecting the output light pulse.
Therefore, $\tau_j$ is the measurement time for the phonon mode
$a_j$ after the collision. Therefore, as long as $\tau_j \ll 1/N$,
the collision events can be measured individually.

 Moreover, to make
sure the success of the output field detection, the phonon number
after the collision requires to be added by more than one. For the
first case, we suppose the collision is completely elastic. Parts of
the molecular movement, which is perpendicular to the surface of the
collision point, will change in direction after the collision
\cite{PhysRev.23.710}.
The average increase of the phonon number for $a_j$ is
$ n_{j0}=2m_m^2 \langle v_j^2\rangle/(\hbar \omega_j m)$ with
$\langle v^2_j\rangle$ the the mean velocity square along the axis
$q_j$. As a result, the requirement for the phonon number change
could be rewritten as $2 k_BT_{\mathrm{env}}> \hbar\omega_j
(m/m_m)$. If the collision is completely inelastic, the molecule
will attach on the surface of the nanosphere for a while before
being kicked out \cite{PhysRev.23.710}. The output velocity
distribution is completely determined by the temperature of the
nanosphere surface. The criteria should be either
$k_BT_{\mathrm{env}}> 2\hbar\omega_j (m/m_m)$, or
$k_BT_{\mathrm{sur}}> 2\hbar\omega_j (m/m_m)$, where
$T_{\mathrm{sur}}$ is the temperature of the surface of the
nanosphere. To distinguish elastic and inelastic collision, we can
cool the temperature to the limit that $k_BT_{\mathrm{env}}\ll
\hbar\omega_j (m/m_m)$, and makes the condition
$k_BT_{\mathrm{sur}}> 2\hbar\omega_j (m/m_m)$ fulfills by adding a
long wavelength laser to heat the sphere.  If the collisions are all
elastic, there is no signal on the photon detectors. If there are
parts of the collisions are inelastic, there are output pulses of
lights. Besides, the distribution of the photon numbers is
determined by the surface temperature of the sphere. In other words,
we can measure the surface temperature of the nanosphere by
detecting the output light pulses.

Besides, if there are more than one type of molecules involved, we
can also distinguish them by the measurement.
We suppose that the energy increasing
of the phonon mode $a_j$ after collision fulfills the Maxwell-Boltzmann distribution and the
collisions are elastic.
The mean phonon increasing for mode $a_j$ is $n_{j0} = 1/(e^{\hbar\omega_j/k_B T_j}-1)=
 2k_B T_{\mathrm{env}} m_a/(\hbar \omega_j m)$, where $T_j$ is the effective temperature
of mode $a_j$ after single collisions. The phonon adding distribution after single collisions
for mode $a_j$ is $f(n_j)dn =\frac{2}{\sqrt{\pi}} (\frac{\hbar \omega_j}{k_B T})^{3/2} \exp (-n \frac{\hbar \omega_j
}{k_B T}) dn$ \cite{Lau05}. However, the mean phonon adding cannot be measured from a single light
pulse. As the photon detector can only measure the
photon pulses number in integer. The measured number distribution of the mode $a_j$ should be Bose-Einstein distribution, which is $f(n_j) = n_{j0}^{n_j} /(1+n_{j0})^{n_j+1}$ \cite{QO97}.
We suppose the mass
$m_{a}$ and $m_{b}$ corresponding to the molecules $a$ and $b$,
respectively, and the same mean kinetic energies for both types of
the molecules. The average increase of the phonon number for the
phonon modes $a_j$ is different for different collision.
As shown in Fig. 2a, there are two curves in the phonons distribution of mode $a_{3}$,
which represent the two different molecules. Fig. 2b shows The measured phonon distribution
for different molecules. We can distinguish the different molecules from data fitting.

Specifically, we consider the example
below. We consider a sphere with radius $r=50$ nm and mass
$m=1.03\times 10^{-18}$ kg ($\rho= 1.96 \mathrm{g/cm}^3$). The optical
 tweezer is consturcted with a laser with power $P_t= 25$mW at wavelength
 $\lambda=1500$nm, and a lense of numerical aperture $N=0.9$. The trap
frequency is $(\omega_1,\omega_2,\omega_3)/2\pi \simeq (0.5,0.5,0.2)$
MHz \cite{2010arXiv1010.3109R}. We consider the cavity with length $L=5$ mm, mode waist $\omega =
10\mu$m and wavelength $1.5\mu$m. In the case of $\epsilon \gg 1$,
we have $g_z= 52.2$ Hz, $g_x=g_y=2.2$ Hz, and the zero point
fluctuation $(z,x,y)_{\mathrm{zpf}} = \sqrt{\hbar/(2mw_j)}= 
(4.0,4.0,6.4) \times 10^{-12}$ m. In order to have the final phonon
number $n_{1,2,3}< 1$, the finesse of the cavity should be around
$10^5$. For $m_m = 6.63 \times 10^{-26}$kg and the gas pressure
$10^{-10}$Torr, the collision events per second are about $10$. If
we suppose the cavity decay rate to be $\kappa = 0.5$ MHz,
corresponding to finesse $2\times10^5$, with proper driving strength
($|\alpha_1|^2 \sim 5\times10^4$, $|\alpha_2|^2\simeq 2.5\times 10^7$,
$|\alpha_3|^2\sim
10^7$), the cooling rate for all three modes would be $10^2$Hz and
the mean addition of the phonon number for $a_j$ after each
collision is around $4$. Therefore individual measurements for the
collision events can be distinguished for the three phonon modes.
The cooling laser power for cavity mode $a_{c1}$ is in the order of $10^{-8}$W,
The laser powers for cavity modes $a_{c2}$ and $a_{c3}$ are in the order of $10^{-5}$W.
\begin{figure}[htbp]
  \centering
  \includegraphics[width=9cm]{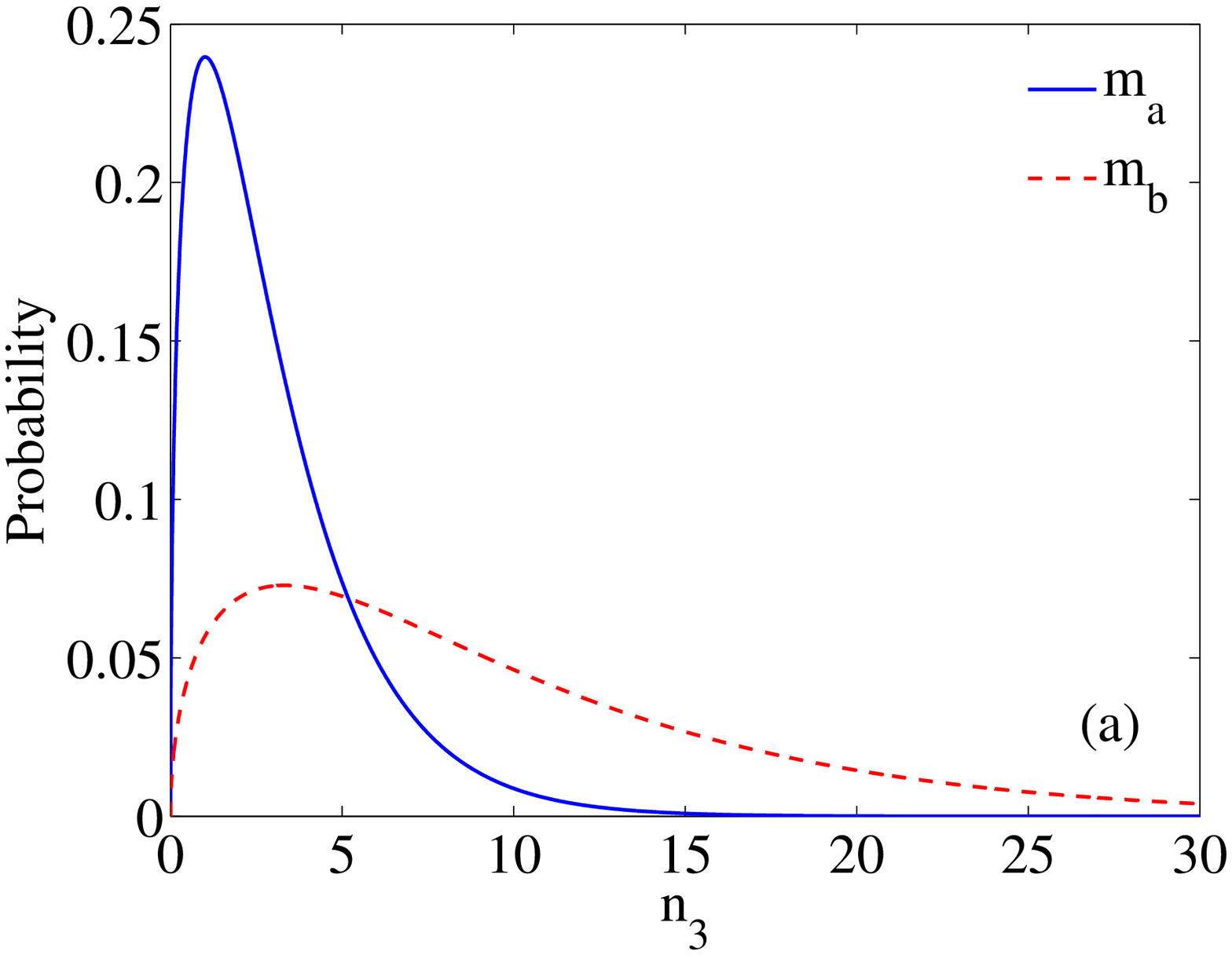}
  \includegraphics[width=9cm]{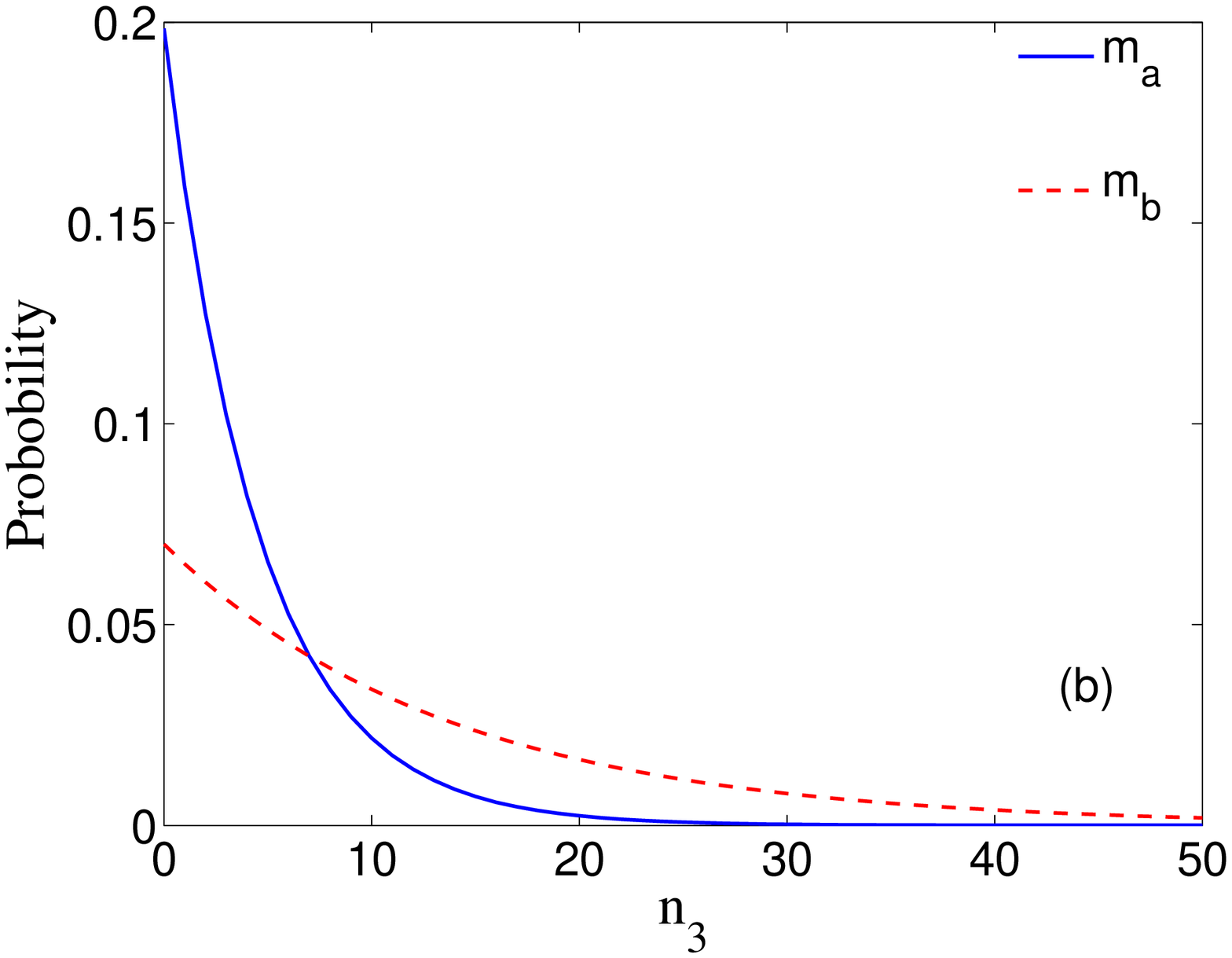}
\caption{(color online) (a) Distribution of the mean phonon increase, (b) measured phonon number
increase distribution of a
mechanical mode $a_3$ after a elastic collision between the nanosphere and a
molecule with mass $m_a=6.63\times 10^{-26}$ kg or
$m_b=2.18 \times 10^{-25}$ kg. The temperature of the gas is $300$K.}
\end{figure}

So far we haven't considered the systematic noise effects in our
treatment. In real experiments, however, the noise from lasers may
be fatal to the success of an experiment. We first consider the
heating effects from the optical trap \cite{PhysRevA.56.R1095}. If
we want to realize motional ground state cooling of the nanosphere,
the heating rate should be much smaller than the laser cooling rate.
Strictly speaking, the heating comes from the laser intensity
fluctuation and the laser-beam-pointing noise. For the former, we
define the fluctuations of the laser $\epsilon(t) = (I(t)-I_0)/I_0$,
with $I_0$ the average intensity and $I(t)$ the laser intensity at
time $t$. By using first-order time-dependent perturbation theory,
we get  $\langle \dot{E}\rangle = \frac{\pi}{2} \omega_j^2
S_\epsilon (2 \omega_j) \langle E\rangle$ \cite{PhysRevA.56.R1095}.
The heating constant is $\Gamma_\epsilon = \frac{\pi}{2} \omega_j^2
S_\epsilon (2\omega_j)$, where $S_\epsilon (\omega) = \frac{2}{\pi}
\int^\infty_0 d\tau \cos (\omega\tau) \langle \epsilon(t)
\epsilon(t+\tau)\rangle$ is the one-sided power spectrum of the
fractional intensity noise, which could be on the order of
$10^{-14}\mathrm{Hz}^{-1}$. For the trap frequency of MHz,
$\Gamma_\epsilon$ approaches the order of $10^{-1}$Hz. The
laser-beam-pointing noise is originated from the fluctuation
relevant to the location of the trap center, which is independent of
the phonon energy. Similarly, we may get $\langle\dot{E} \rangle =
\frac{\pi}{2} m \omega_j^4 S_j (\omega_j),$ where $j=x,y,z$, and
$S_j(\omega)$ is the noise spectrum of location fluctuations. We
define the heating rate as $ \Gamma_j = \frac{\pi}{2} m \omega_j^4
S_j (\omega_j) /(\hbar \omega_j)$, which represents phonon number
increase per second. If we set $\Gamma_j$ to be on the order of
$10^{-1}$Hz, we should make sure that $S_j (\omega_j)$ is around
$10^{-35}\mathrm{m}^2/$Hz for $\omega_j\sim 1$MHz. Experimentally
$S_j(\omega)$ has been controlled less than $10^{-34} \mathrm{m}^2/$
Hz for $\omega \sim 2\pi$ kHz \cite{PhysRevLett.99.160801}. With the
increase of the optical trap frequency to large detuning from the
system's resonant frequency, $S_j(\omega_j)$ is dropping down
quickly. Therefore, we believe that the laser-beam-pointing noise
could be well controlled and the heating rate $\Gamma_j$ would be
less than $0.1$ Hz.

The phase noise induced by the cooling laser also need to be
seriously considered
\cite{PhysRevA.78.021801,PhysRevA.80.063819,Yin2009}. Because the
cooling laser is of finite linewidth, the laser field can be wrote
down as $\varepsilon (t) = \varepsilon e^{i\phi(t)}$. We assume the
phase noise $\phi(t)$ to be Gaussian and with zero mean value. For
the Lorentzian noise spectrum with $S_{\dot{\phi}}(\omega) =
2\Gamma_L \gamma_c/(\gamma_c^2 + \omega^2)$, and correlation
function $\{ \dot{\phi(s)} \dot{\phi(s')} \} = \Gamma_L \gamma_c
\exp (-\gamma_c| s-s'|)$, where $\Gamma_L$ is the linewidth of the
laser and $\gamma_c^{-1}$ is the correlation time of the laser phase
noise, the phonon number limited by this noise is
 $n_{ph} > n_c \frac{\Gamma_L}{\kappa}
\frac{\gamma_c^2}{\gamma_c^2+ \omega_j^2}$
\cite{PhysRevA.80.063819}. If we choose $\Gamma_L =1$ kHz, $\gamma_c
= 3$ kHz, $\omega_j=10^6$ Hz, and $n_c=10^7$,  we have $n_{ph} \ll
1$. In this sense, like above discussed noise effects, the phase
noise effect can also be neglected.

\section{conclusion} \label{sec:c}

In conclusion, we have proposed a scheme to cool and measure the 3D
motion of an optically trapped nanosphere confined in a single
cavity, driven by three lasers. With properly locating the optical
trap and the laser detunings, we have shown by calculation that the
3D motion of the nanosphere could be cooled and detected
simultaneously, and down to ground states if the sideband resolved
condition is fulfilled. We have justified the experimental
feasibility of our scheme under currently available technology. We
argue that our scheme would be useful for not only checking the
Maxwell-Boltzmann distribution at single-collision level, but also
measuring the temperature of the surface of the nanosphere and the
mass of the molecule.


The work is supported by NNSFC under Grant No. 10974225, by CAS and
by NFRPC. TL would like to thank M. G. Raizen for helpful
discussions and the suggestion of using a cooled bead to detect
single molecules.

{\em Note added}: After submitting the paper, we have found a
related experimental paper \cite{2010arXiv1012.2156Z}, which
eliminates degenerate trajectory of single atom strongly coupled to
the tilted cavity TEM10 mode.


\end{document}